\documentclass{PoS}

\usepackage{graphicx}              
\usepackage{amsmath}               
\usepackage{amsfonts}              
\usepackage{amsthm}   		         
\usepackage{color}	
\usepackage{psfrag}
\usepackage[all, knot]{xy}
\usepackage{calc}

\newtheorem{theorem}{Theorem}

\newcommand\nn{\nonumber}
\newcommand\be{\begin{eqnarray}}
\newcommand\ee{\end{eqnarray}}

\newcommand{\cC}{{\mathcal C}}

\newcommand{\cF}{{\mathcal F}}

\newcommand{\cH}{{\mathcal H}}

\newcommand{\cK}{{\mathcal K}}

\newcommand{\cM}{{\mathcal M}}

\newcommand{\cP}{{\mathcal P}}

\newcommand{\cS}{{\mathcal S}}
\newcommand{\cT}{{\mathcal T}}

\newcommand{\fc}{\mathfrak{c}}

\newcommand\U{{\mathrm U}}
\newcommand\C{{\mathbb C}}
\newcommand\N{{\mathbb N}}
\newcommand\R{{\mathbb R}}

\newcommand\T{{\mathbb T}}

\newcommand{\SO}{{\rm SO}}
\newcommand{\SL}{{\rm SL}}
\newcommand{\SU}{{\rm SU}}

\newcommand{\Sl}{\mathfrak{sl}}
\newcommand{\su}{\mathfrak{su}}

\newcommand{\braaket}[2]{\langle #1  |   #2 \rangle }

\newcommand{\braaaket}[3]{\langle #1  |   #2  |  #3 \rangle }
\newcommand{\bra}[1]{\langle #1  |}
\newcommand{\ket}[1]{| #1 \rangle}
\newcommand{\ketr}[1]{| #1 ]}
\newcommand{\brar}[1]{[ #1 |}
\newcommand{\la}{\langle}
\newcommand{\ra}{\rangle}

\newcommand\tX{\tl X}

\newcommand{\tz}{\tilde{z}}

\newcommand{\twovec}[2]{\begin{pmatrix} #1 \\ #2 \end{pmatrix}}

\newcommand{\tl}{\widetilde}
\newcommand{\f}{\frac}

\newcommand{\id}{\mathbb{I}}

\title{Spinors and Twistors in Loop Gravity \\ and Spin Foams}
\ShortTitle{Spinors and Twistors in LQG and SF}

\FullConference{3rd Quantum Gravity and Quantum Geometry School,\\
		February 28 - March 13, 2011\\
		Zakopane, Poland}

\author{Maite Dupuis$^{a,c}$, \speaker{Simone Speziale}$^b$ and Johannes Tambornino$^a$\\
\llap{$^a$}Laboratoire de Physique, ENS Lyon, CNRS-UMR 5672, 46 All\'ee d'Italie, Lyon 69007, France\\
\llap{$^b$}Centre de Physique Th\'eorique, CNRS-UMR 6207, Luminy Case 907, 13288 Marseille, France\\
\llap{$^c$}Institute for Theoretical Physics III, University of Erlangen-N\"urnberg, Erlangen, Germany\\
}

\PoS{PoS(QGQGS 2011)015}

\abstract{Spinorial tools have recently come back to fashion in loop gravity and spin foams. They provide an elegant tool relating the standard holonomy-flux algebra to the twisted geometry picture of the classical phase space on a fixed graph, and to twistors.
In these lectures we provide a brief and technical introduction to the formalism and some of its applications.
}

\begin{document}

\section{Introduction}

The Hilbert space  $\cH_{{\rm LQG}}$ used in loop quantum gravity can be heuristically understood as a collection of certain Hilbert spaces associated 
to all possible graphs.\footnote{ In loop gravity the Hilbert space associated to a graph $\Gamma$ is given by $\cH_\Gamma := L^2(\SU(2)^E, d^Eg)$ where $E$ is the number of edges of that graph and $d^Eg$ the product Haar measure.
The Hilbert space of the continuum theory arises from the individual graph-Hilbert spaces as an inductive limit
 $
 \cH_{\rm LQG} := \overline{ \cup_\Gamma \cH_\Gamma / \sim   }
 $,
 where $\sim$ denotes an equivalence relation between states living on different graphs and the completion in an appropriate topology is taken.}
While being big enough to represent the infinite number of degrees of freedom of the gravitational field, the space is made of simple building blocks, the finite-dimensional Hilbert spaces $\cH_\Gamma = L^2(\SU(2)^E, d^Eg)$, where $E$ is the number of edges of that graph and $d^Eg$ the product Haar measure.
Each graph space carries only a finite number of degrees of freedom, therefore $\cH_\Gamma \subset \cH_{\rm LQG}$ 
corresponds to a  truncation of the full theory \cite{Rovelli:2010km}.
The resulting finite degrees of freedom are usually interpreted in terms of continuous but distributional configurations of the gravitational field \cite{Ashtekar:1998ak, Thiemann:2000bv}.\footnote{An alternative but analogously continuous interpretation has been proposed in \cite{Bianchi:2009tj,Freidel:2011ue}. See also \cite{Baratin:2010nn} for a related Hilbert space representation based on the non-commutative flux variables.}
On the other hand, it has been shown in \cite{Freidel:2010aq,Freidel:2010bw} and then \cite{Rovelli:2010km,Bianchi:2010gc} that the same degrees of freedom describe a new discretized version of general relativity, more general than Regge calculus, called \emph{twisted geometries}.
These are discrete geometries where each building block is described by an elementary polyhedron; however, the polyhedra are glued together in a ``twisted'' way such that the resulting metric is discontinuous at the faces. Spin network functions, which form an orthogonal basis of each $\cH_\Gamma$, can therefore be interpreted as the quantum versions of these polyhedral geometries \cite{Bianchi:2010gc}.
Related ideas have been pushed further in \cite{Freidel:2009ck, Freidel:2010tt, Borja:2010rc}, and unraveled a powerful $\U(N)$ symmetry in the intertwiner space.

Remarkably, all these structures are captured in a simple and elegant way using spinors. This is rooted in the elementary fact that Lie groups carry a complex structure which can be used to give a representation of the algebra in terms of harmonic oscillators. The fundamental space $\cH_\Gamma$, and the classical phase space behind it, can be described using spinors and twistors. This is done associating a spinors to each half-edge of the graph, and leads to the usual variables, a group element and a Lie-algebra element on each edge. In these short lectures, we give a concise introduction to the formalism, and some of its applications.

\section{Twisted geometries, spinors and null twistors} \label{sec:twisted_geo}

The non-gauge-invariant Hilbert space $\cH_\Gamma$ is a tensor product of individual spaces associated to each edge $e$, $\cH_e = L^2(\SU(2))$. 
Hence, for the moment we restrict attention to a single edge. The space carries a representation of the angular momentum algebra as derivative operators, and the matrix elements 
of SU(2) group elements as multiplicative operators. The total algebra is known as holonomy-flux algebra, from its interpretation in terms of Ashtekar variables.
The space is a quantum version of the phase space  $T^*\SU(2)$, the cotangent bundle of $\SU(2)$, with its canonical Poisson algebra.
The bundle trivializes as $T^*\SU(2) \simeq \SU(2) \times \su(2)$. Accordingly, it can be parameterized using coordinates $(g,X)\in \SU(2)\times \su(2)$, with $g$ being the holonomy of the Ashtekar connection $A$ along the edge $e$, and $X$ the flux of the densitized triad $E$ along a surface dual to the edge and infinitesimally near its initial point.

This structure can be easily obtained from the simpler space $\T = \C^2\times\C^2$ with canonical Poisson brackets \cite{Freidel:2010bw}. One
considers two spinors, say $\ket{z}$ and $\ket{\tz}$, living respectively at the initial and final vertex of $e$.\footnote{Recall that $\Gamma$ is an oriented graph.
Our notation is as follows: 
$\sigma_i$ are the Pauli matrices, $\tau_i = -i/2 \sigma_i$ the Hermitian generators. 
A spinor $\ket{z} \in \C^2$ has components ${z^0},{z^1}$, 
and a dual $\ketr{z} := \epsilon \ket{\bar{z}}$, where $\epsilon=-i\sigma_2$. 
The inner product on $\C^2$ is denoted by $\bra{z}{w}\ra := \bar{z}^0 w^0 + \bar{z}^1w^1$. Dirac's notation is adopted to avoid explicit indices, however the reader should keep in mind that we are dealing here with \emph{classical} spinors, and not quantum states.}
Each spinor is equipped with the canonical symplectic structure 
\be\label{zz}
\{ \ket{z} , \bra{z} \} = -i\ \id.
\ee
We construct a vector in $\R^3$ by projection onto the Pauli matrices,
\be \label{vector}
\vec{X}(z) := \f12 \bra{z}{\vec{\sigma}}\ket{z},
\ee
invertible up to a phase: $\ket{z}$ and $e^{i\varphi}\ket{z}$ define the same vector. Using \eqref{zz}, it is easy to see that the vector carries a Poisson representation of the Lie algebra $\su(2)$,
\be
\{ X^i(z) , X^j(z)  \} = \epsilon^{ijk} X^k(z).
\ee
Similarly, we obtain an independent representation of the algebra using the second spinor, $\ket{\tz}$, which we denote as $\tl X(\tz)$, or $\tX$ in short.
Then, we define the matrix \cite{Freidel:2010tt} 
\be \label{g}
g(z, \tz) := \frac{\ket{z}\brar{\tz} - \ketr{z}\bra{\tz} }{\sqrt{\braaket{z}{z}\braaket{\tz}{\tz}}} ,
\ee
which is well-defined provided the norm of both spinors is non-zero. Such singular configurations can be safely excluded, as we explain below, and in the following we will often restrict attention to the space $\C^2_\ast:=\C^2-\{\bra{z}{z}\ra=0\}$.
Notice that  $g \ket{\tz} = - \alpha \ketr{z}$, where the proportionality coefficient $\alpha$ is the ratio of the spinor norms. 

Consider then the following \emph{area matching} constraint in $\T_\star$,
\be \label{cM}
\cM := \braaket{z}{z} - \braaket{\tz}{\tz} = 0.
\ee
The constraint ensures that the two vectors have the same norm, $|\vec{X}| \stackrel{!}{=} |\vec{\tilde{X}}|$, and that $g$ is unitary and of unit determinant.
It generates $\U(1)$ transformations on the spinors,\footnote{The finite action is obtained via the exponential map,
\[
e^{\{ \theta \cM, \cdot\} } \ket{z} := \sum_{n}\f{\theta^n}{n!} \{\cM, \ket{z}\}_{(n)} = e^{i\theta} \ket{z}.
\]
}
\be
\ket{z} \stackrel{\U(1)}{\rightarrow} e^{i\theta}\ket{z}, \qquad \ket{\tz} \stackrel{\U(1)}{\rightarrow} e^{-i\theta} \ket{\tz},
\ee
that leave invariant $X, \tilde{X}$ as well as the group element $g$. Hence, they provide coordinates for the six-dimensional phase space obtained by symplectic reduction
$\T_\ast/\!/\cM$.
Furthermore, the Poisson algebra induced on the constraint hypersurface $\cM = 0$ is the canonical one of $T^*\SU(2)$, that is
\be\label{Ptsu2}
\begin{tabular}{lll}
$\{X_i, X_j \} = \epsilon_{ijk} \, X_k,$
& $\{\tX_i, \tX_j \} = \epsilon^{ij}{}_k\, \tX_k, $
& $\{X_i, \tX_j \} = 0,$ \\ &&\\
$\{X_i, g\} = -\tau_i g,$ & $ \{\tX_i, g \} = g\tau_i$, & $ \{g_{AB}, g_{CD} \} = 0.$
\end{tabular}
\ee
The two vectors ${X}(z)$ and ${\tilde{X}}(\tz)$ turn out to be related via the adjoint representation,
\be
\tilde{X}(\tz) = -g^{-1}(z,\tz) X(z) g(z,\tz).
\ee
That is, $X$ and $\tl X$ are the alternative parameterizations of $T^*SU(2)$ by right- and left-invariant vector fields.
The results are summarized by the following theorem.

\begin{theorem} \label{thm1} \cite{Freidel:2010bw}
The symplectic reduction of the space $\C^2_\ast \times \C^2_\ast$ by the constraint $\cM$ is isomorphic to the cotangent bundle $T^*\SU(2) - \{ |X| = 0  \}$ as a symplectic space.
\end{theorem} 

Finally, the isomorphism can be extended to the full space $T^*SU(2)$ by suitable reduction on the singular configurations of zero norm, which ensures that one recovers the right topology.\footnote{At the singular points, we have the following two different spaces: on the spinor side, $\T$ such that $\la z\ket{z}=\la \tz\ket{\tz}=0$ describes the manifold $S^1\times S^2\times S^2$; on the SU(2) side, $|X|=0$ describes the manifold $S^3 \cong S^1\times S^2$. The reduction needed to include the singular configurations consists simply of identifying the two $S^2$ of the spinor side. See  \cite{Freidel:2010aq} for details.}
The theorem provides a classical counterpart to the well-known Schwinger representation of the quantum angular momentum.
It shows that the Hilbert space $\cH_e$, associated to a single edge in loop gravity, can be understood as the quantization of (a certain symplectic reduction of) the classical phase space spanned by two spinors, interpreted as living on the initial and final vertex respectively. 

\subsection{Twisted geometries}
The spinorial description is related to twisted geometries in a simple way. Recall \cite{Freidel:2010aq} that twisted geometries parameterize $T^*\SU(2)$ in terms of a conjugated pair $(\lambda,\xi)\in \R^+\times S^1$ and two unit vectors $N(\zeta)$ and $\tl N(\tl \zeta)$, where $\zeta$ and $\tl \zeta$ are stereographic complex coordinates on $S^2$. Then, the relation to spinors is given by the map
\be\label{twigeo}
\lambda=\f{\la z|z\ra}2,\qquad
\xi = -2\arg(z^1)-2\arg(\tz^1), \qquad
\zeta=\f{z^0}{z^1},\qquad
\tl \zeta=\f{\tz^0}{\tz^1}.
\ee
The last two equations can be recognized as Hopf projections $S^3\to S^2$.
Using \eqref{vector}, we immediately get $X(z) \equiv \lambda N(\zeta)$.
In terms of these variables, the Poisson algebra \eqref{Ptsu2} reads
\be\label{PP}
\begin{tabular}{lll}
$\{X_i, X_j \} = \epsilon_{ijk} \, X_k,$
& $\{\tilde X_i, \tilde X_j \} = \epsilon^{ij}{}_k\, \tilde X_k, $
& $\{X_i, \tilde X_j \} = 0,$ \\ &&\\
$\{\lambda, \xi\} = 1$, & $\{\xi, X_i \} = L_i(\zeta)$, & $\{\xi, \tilde X_i \} = L_i(\tilde \zeta)$,
\end{tabular}
\ee
where
\be
L(\zeta) = (-\bar\zeta,-\zeta,1)
\ee
is a connection for the Hopf bundle compatible with the Hopf section. As anticipated, $\lambda$ and $\xi$ are conjugate variables.

\subsection{Twistors}

The above structure can be also conveniently described in terms of a twistor  \cite{Freidel:2010bw}. 
For our purposes, a twistor is simply a pair of spinors transforming under the defining representation of the Lorentz group $\SO(3,1)$. The reader should be familiar with the fact that the phase spaces $\C^2$ and $\T$ carry also a natural action of $\SL(2,\C)$.\footnote{In the spinorial representations, the boosts are given respectively 
by $(1/2) {\rm Re} [z|\sigma\ket{z}$ and ${\rm Re} [\tz|\sigma\ket{z}$ \cite{Dupuis:2011wy}.} However, the form \eqref{zz} of the Poisson brackets is not invariant under the boosts, which is an immediate consequence of the absence of an $\SL(2,\C)$ positive-definite scalar product. An invariant symplectic structure can be obtained through a linear transformation to new spinors with a well-defined handedness,
\be
\ket{r} := \frac{1}{\sqrt{2}} \Big(\ket{z} + i \ketr{\tz} \Big), \qquad \ket{l} := \frac{1}{\sqrt{2}} \Big(\ket{z} - i |\tz] \Big).
\ee
In the new variables, the Poisson brackets \eqref{zz} read\footnote{We adopt here the standard spinor conventions of distinguishing right-handed indices with a dot.}
\be \label{ut}
\{ \ket{r} , \bra{l}  \} = -i\, \id.
\ee
Furthermore, the quantities  
\be\label{defJK}
\vec{J} = {\rm Re} \braaaket{l}{\vec{\sigma}}{r}, \qquad \vec{K} = {\rm Im} \braaaket{l}{\vec{\sigma}}{r},
\ee
generate an $\SL(2,\C)$ algebra, and their action leaves \eqref{ut} invariant \cite{Dupuis:2011wy}.

The twistor $Z = (\ket{r}, \ket{l}) \in\T$ transforms under $G\in \SL(2,\C)$ as
\be
G \, Z =
\twovec{G\, \ket{r}}{  (G^\dagger)^{-1}\, \ket{l}},
\ee
and it comes with a natural, $\SL(2,\C)$-invariant (non positive definite) bilinear form,
\be
({Z'},Z) := \braaket{{l'}}{r} + \braaket{{r'}}{l}.
\ee
Using this we can define the \emph{helicity} of a twistor as $s(Z) := \frac{1}{2}(Z,Z)$. 

Reformulated in these new variables, the constraint $\cM$ can be seen to impose the vanishing of the helicity:
$\cM(l,r) = 2\, s(Z)$. Hence, the area matching means restricting attention to \emph{null} (i.e. zero helicity) twistors, and $\cM$ generates a $\U(1)$ transformation,
\be
Z \stackrel{\U(1)}{\mapsto} e^{i\varphi} Z.
\ee
The result of the previous section can be now summarized as a parameterization of $T^*\SU(2)$ as the canonical phase space of null twistors up to U(1) transformations.
In this picture, we do not distinguish between source and target of the edge:
now there is simply \emph{one twistor} associated to each edge.

\section{Gauge-invariant spinor networks and polyhedral geometries}\label{sec:closure}

Let us now generalize the previous results from a single edge to the full oriented graph $\Gamma$: two spinors are then living on each edge. We adapt the notation by removing the tildes, and denoting explicitly the spinors in terms of source and target nodes, $z_e^s$ and $z_e^t$. 
Local gauge invariance is imposed at each node of the graph by the closure condition,
\be\label{closure}
\fc^n=\sum_{e\in n} X_e = \sum_{e\in n} \lambda_e N_e = 0.
\ee 
On each node, \eqref{closure} gives three first class constraints, generating SU(2) transformations.
Denote $ S_\Gamma=P_\Gamma/\!/ \fc^N$ the reduced space, where $P_\Gamma= (\C_\ast^2\times \C_\ast^2 /\!/\cM)^E\cong T^*\SU(2)^E$.
The result is the gauge-invariant phase space\footnote{More precisely, a symplectic manifold itself up to singular points, see for example \cite{Bahr:2007xn}.} of loop gravity on a fixed graph.
That is, $ S_\Gamma = T^*\SU(2)^E/T^*\SU(2)^N$ \cite{Freidel:2010aq}. Gauge invariance eliminates the simple Cartesian structure of the initial phase space.

In terms of spinor, \eqref{closure} is equivalent to the $2\times2$ matrix equation 
\be\label{clos}
\fc^n:=\sum_e |z^n_e\ra \la z^n_e | - \f12 \sum_e \la z^n_e | z^n_e \ra \id=0.
\ee
The spinorial description of the symplectic structure on $ S_\Gamma$ is summarized by the following action, with $\tau$ an auxiliary variable playing the role of time,
\be\label{spinornet}
\cS_\Gamma[ z^{s,t}_e]
\,\equiv\,
\int d\tau\,
\sum_e -i\la  z^{s,t}_e|\partial_\tau z^{s,t}_e\ra
+\sum_e \Phi_e(\la  z^{s}_e| z^{s}_e\ra-\la  z^{t}_e| z^{t}_e\ra)
+\sum_n \sum_{e\ni n} \la  z^n_e|\Theta_n| z^n_e\ra.
\ee
The scalars $\Phi_e$ and the traceless matrices $\Theta_n$ are the Lagrange multipliers for area-matching and closure constraints respectively. 
This quantity defines a notion of \emph{spinor network} \cite{Borja:2010rc}, as the classical counterpart of a spin network. 
This should not be confused with a coherent spin network state. 
What we mean is that quantizing the classical Poissonian structure described by \eqref{spinornet}, we end up with the Hilbert space of gauge-invariant spin networks equipped with a representation of the holonomy-flux algebra. Conversely, any family of coherent spin network states is labelled by a point in such spinorial phase space.\footnote{Different families of coherent spin networks exist, which differ in their peakedness properties. Regardless of their specific form, they can always be labeled by a point in that phase space.}

The reduced phase space $ S_\Gamma$ can be further given a geometric interpretation in terms of polyhedra, using the twisted geometry variables \eqref{twigeo}.
Let us briefly review this useful result, as an application of the formalism.
Consider first a single node, say of valence $F$. Associated with it, there is a set of $F$ unit vectors $N_e$, $e=1 \ldots F$, plus a norm, $\lambda_e$,
and an angle, $\xi_e$. 
An old theorem by Minkowski guarantees that if \eqref{closure} holds, these data reconstruct a unique polyhedron dual to the node:

\begin{theorem} \label{thm2} [Minkowski]
If $N_1,\ldots,N_F$ are non-coplanar unit vectors and $\lambda_1,\ldots,\lambda_F$ are positive numbers such that the closure condition  
\eqref{closure} holds, than there exists a bounded, convex polyhedron whose faces have outwards normals $N_e$ and areas $\lambda_e$, unique up to rotations and translations. 
\end{theorem} 

See \cite{Bianchi:2010gc} for more details and the explicit algorithm for the reconstruction of the polyhedron.
Thanks to the theorem, a spinor network can be interpreted as a collection of polyhedra with adjacency relations established by the connectivity of the graph.
The emerging geometric picture has an important peculiarity, that motivates its name of \emph{twisted} geometries.\footnote{Of course, the existence of a description in terms of twistor is also a reason for the name!}
In fact, the geometries are piecewise flat, but can be discontinuous at the faces connecting the polyhedra.
The origin of the discontinuity lies in the fact that the local face geometry of a polyhedron depends on the entire set of data of the polyhedron.
A face shared by two adjacent polyhedra has, by definition, the same area, but there is nothing that guarantees the same shape. Hence, the metric on a face jumps when switching from one frame to the next, in particular the lengths, angles and even number of sides of a face change discontinuously.

To make the geometries continuous, one needs to include additional \emph{shape-matching} conditions. For the special case of a four-valent graph, dual to a triangulation, these were found in \cite{Dittrich:2008va},
and effectively reduce the data to edge lengths, thus Regge calculus is recovered. For general graphs,
they were studied in \cite{Bianchi:2010gc}, and the result would be a generalization of Regge calculus to arbitrary 3d cellular decompositions.\footnote{Note that the fundamental variables of such a generalization would be areas and angles, and not edge lengths.}
Such discontinuity might appear appalling at first, but it can be argued for: after all, standard Regge calculus is torsion-free, whereas the kinematical phase space of loop quantum gravity should carry room for torsion. 
In this geometric picture, the angles $\xi_e$ carry a notion of discrete \emph{extrinsic} geometry among the polyhedra. The presence of extrinsic geometry is granted by the symplectomorphism with the algebra of $T^*SU(2)$, which is related to a discretization of the phase space of general relativity. 
An elegant result, in this perspective, is the \emph{abelianization} of part of the Poisson algebra \eqref{PP}, with the canonically conjugated $\lambda_e$ and $\xi_e$ capturing some ``scalar'' components of intrinsic and extrinsic geometry.
See \cite{Freidel:2010aq} for more details on the relation between $\xi_e$ and extrinsic curvature.
In Regge geometries, the extrinsic curvature is also captured by an angle, the dihedral angle between tetrahedra. 
$\xi_e$ can be similarly thought of, although a naive identification with a dihedral angle is prevented by the discontinuous nature of the twisted geometries.\footnote{In fact, it has been shown in \cite{Dittrich:2008ar} that prior to imposing the shape matching conditions in a triangulation, there are three independent dihedral angles per triangle, one associated with each side of the triangle. It is only when the shapes match that the three coincide, and the usual dihedral angle emerges.}

\subsection{$\U(N)$ formalism}
The spinorial description of $ S_\Gamma$, captured in \eqref{spinornet}, unravels also the existence of a $\U(N)$ symmetry acting on the nodes.
This symmetry arises if we allow for the individual areas -- the norms $\lambda_e$ -- to vary, but keeping the value of their sum around each node fixed,
and it is the classical version of the $\U(N)$ symmetry for the space of intertwiners introduced a while back in \cite{Girelli:2005ii}.
To bring to light this symmetry, we use a beautiful property of the spinorial description, which allows us to find a genuine algebra of invariants.
Recall in fact that the basis of invariants commonly used, the scalar products $\vec{X}_e \cdot \vec{X}_f$ ($e$ and $f$ being edges sharing the same node), 
fails to give a Poisson algebra, because of the well-known relation
\be\label{JJJ}
\{\vec{X}_e\cdot \vec{X}_f, \vec{X}_e\cdot \vec{X}_g \} =  \vec{X}_e \cdot \vec{X}_f \times \vec{X}_g.
\ee
A different basis is then required. Let us for the moment ignore the edge constraints $\cM_e$:
Then, the spinor variables completely factorize to sets of independent spinors around each node, as many as the valency of the node. A complete family of quadratic spinor invariants can be identified \cite{Borja:2010rc},
\be\label{EF}
E_{ef}\equiv \la z_e|z_f\ra,\qquad
F_{ef}\equiv [z_e |z_f\ra.
\ee
The matrix $E$ is Hermitian, $E_{ef}=\overline{E}_{fe}$, while the matrix $F$ is holomorphic in the spinor variables and anti-symmetric, $F_{ef}=-F_{fe}$.
It is easy to check that they commute with \eqref{clos}, and that they form an (over)complete\footnote{There exist $(F-2)(F-3)$ independent \emph{Pl\"ucker relations} among them, of the form $F_{ij}F_{kl}=F_{ik} F_{jl} -F_{il}F_{jk}$ \cite{Dupuis:2011wy}.} basis of local SU(2) invariants. For instance, the scalar products are expressed via $\vec{X}_e \cdot \vec{X}_f = \f12\left(|E_{ef}|^2-|F_{ef}|^2 \right)$.
Remarkably, the invariants \eqref{EF} form a genuine algebra \cite{Borja:2010rc}. In particular, the $E$ alone generate a $\mathfrak{u}(N)$ subalgebra.
This property of the $\U(N)$ formalism has important applications which have been used at the quantum level (e.g. \cite{Freidel:2009ck, Freidel:2010tt, Borja:2010rc,Borja:2011pd}), where these invariants are promoted to operators acting on the Hilbert space of intertwiner states with fixed total area.

The individual node spaces $\cal U$ represent shapes of framed polyhedra with fixed total area. Finally, reintroducing the area matching conditions among adjacent polyhedra, the various $E,F$ variables around different nodes become related to one another, and the non-local structure of $S_\Gamma$ is recovered.
All this can be summarized by the following diagram:
\begin{center}
\begin{tabular}{ccc}
{Twistor phase space,} $\underset{e}{\times} \C^4$ \qquad & $\longrightarrow$ & \ holonomy-flux phase space, $\underset{e}\times T^*SU(2)$ \\ 
& \emph{area matching} & \\
$\downarrow$ \emph{closure} & & $\downarrow$ \emph{closure} \\
& & \\
$U(N)$ formalism, $\underset{n}\times {\cal U} $ \qquad & $\longrightarrow$ & \ gauge-invariant space,  \\ 
& \emph{area matching} &  (closed) twisted geometries
\end{tabular}
\end{center}
In both paths, we start from a simple Cartesian product phase space, which is preserved by reduction by either the area mathing or gauge invariance. It is the reduction by  \emph{both} conditions that makes the final space non-trivial.

\section{Spinorial Hilbert space for loop gravity}

We have so far discussed properties of the classical theory. However, the spinorial description has many applications also at the quantum level \cite{Livine:2011gp, Livine:2011zz}. One of the main result is a quantum version of theorem \ref{thm1}, which introduces a coherent spin network representation in terms of the Bargmann holomorphic representation of the harmonic oscillator.
\begin{theorem} \cite{Livine:2011gp}
The $\U(1)$-symmetry reduction of the double copy of the Bargmann space $\cF_2$ of holomorphic, square-integrable functions in two complex variables with respect to a normalized Gaussian measure is unitarily equivalent to the space of square integrable functions over $\SU(2)$ with respect to the Haar measure, $\cF_2 \otimes \cF_2 / \U(1) \simeq L^2(\SU(2), dg)$.  
\end{theorem}
In the remainder of this section we will explain the content and the consequences of this theorem: one surprising feature of the spinorial formalism, which was discussed in \cite{Livine:2011gp}, is that the Haar measure on $\SU(2)$ turns out to be just a Gaussian measure on $\C^4$ when written in terms of spinors, in the sense that 
\be
\int dg f(g) = \int  d\mu(z) \int d\mu(\tz) f(g(z, \tz)), \qquad d\mu(z) := \frac{1}{\pi^2}e^{-\braaket{z}{z}}d^4z \, .\nn
\ee 
for any $f \in L^2(\SU(2))$ and the group element $g$ interpreted as function of spinors as in (\ref{g}) on the right side. Using spinorial variables to characterize $\SU(2)$ can be understood as choosing a coordinate system with a lot of redundant degrees of freedom. Thus, $f(g(z,\tz))$ is constant along certain directions in $\C^4$ which can be used to turn the Haar measure into Gaussian form. We mention also that a similar construction exists for the Haar measure on $\SL(2,\C)$ \cite{Livine:2011vk}.

The usual Hilbert space associated to a single edge in loop gravity is $\cH_e := L^2(\SU(2),dg)$. The symplectomorphism stated in Theorem \ref{thm1} however allows to follow a different route. The most natural space to quantize $\C^2$ and its canonical brackets \eqref{zz} is the Bargmann space $\cF_2$ of square-integrable, holomorphic functions over two complex coordinates with a normalized Gaussian measure, 
\be
\cF_2 := L^2_{\rm hol}(\C^2, d\mu(z)), 
\ee
As there are two spinors, one living on each vertex of $e$, restricted by the $\U(1)$-constraint enforcing them to have equal length, the appropriate space to look for a representation of $T^*\SU(2)$ is
\be
\cH_e^{\rm spin} := \cF_2 \otimes \cF_2 / \U(1) \, ,
\ee
which we call the \emph{spinor Hilbert space} associated to an edge $e$. The spinors $\ket{z}$ and $\ket{\tz}$ are represented on $\cH^{\rm spin}_e$ as ladder operators, $g$ and $X$ are then constructed as composite operators via (\ref{vector}) and (\ref{g}). Restricting attention to $\U(1)$-invariant functions of both spinors singles out polynomials of the form (labeled by $\alpha, \tilde{\alpha} \in \C^2, j \in \frac{1}{2}\N$)
\be \label{basis}
\cP^j_{\alpha \tilde{\alpha}}(z,\tz) := \frac{1}{(2j)!}\braaket{\alpha}{z}^{2j}\brar{\tz}\epsilon \ket{\tilde{\alpha}}^{2j} \, , 
\ee
which are holomorphic in both spinor variables and further have matching degree. They form an over-complete basis of $\cH^{\rm spin}_e$, the completeness relations can be derived as
\be 
\int d\mu(z) \int d\mu(\tz) \overline{\cP^j_{\omega \tilde{\omega}}(z,\tz)} \cP^{k}_{\alpha \tilde{\alpha}}(z,\tz) & = & \delta^{jk}\braaket{\alpha}{\omega}^{2j}\braaket{\tilde{\omega}}{\alpha}^{2j} \, ,\nn \\
\sum\limits_j \int d\mu(\omega) d\mu(\tilde{\omega}) \frac{d_j}{(2j)!} \overline{\cP^j_{\omega \tilde{\omega}}(z_1, \tz_1)} \cP^j_{\omega \tilde{\omega}}(z_2, \tz_2) & = & I_0(2\braaket{z_1}{z_2} \braaket{\tz_1}{\tz_2}) \, . \nn
\ee
Here $I_0(x)$ is the zeroth modified Bessel function of first kind. It plays the role of the delta-distribution on $\cH^{\rm spin}_e$, in the sense that
\be
\int d\mu(z) \int d\mu(\tz) I_0(2\braaket{z}{w}\braaket{\tz}{\tilde{w}}) f(z,\tz) = f(w,\tilde{w}) \qquad \forall f \in \cH^{\rm spin}_e
\ee
The basis elements \eqref{basis} can be extented to the whole graph, and gauge-invariance imposed at the nodes via the usual group-averaging technique. The result is a new type of coherent spin networks, based on the coherent states for the harmonic oscillator.\footnote{See \cite{QTwiGeo} for comparison among some coherent spin network proposals.}
In particular, it carries a holomorphic representation of the holonomy-flux algebra, which is unitarily equivalent to the standard one. To see this, notice that the above completeness relations are, up to a missing factor of $d_j := 2j+1$ on the right side, exactly the ones fulfilled by the Wigner matrix elements in $L^2(\SU(2),dg)$ when written in the coherent state basis. Thus it is immediate to see that the two spaces are unitarily equivalent. The unitary map can explicitly be written in terms of an integral kernel as
\be \label{map}
\cT_e: && \cH_e \rightarrow \cH^{\rm spin}_e; \\
     && f(g) \mapsto (\cT_e f)(z, \tz) := \int dg \cK_g(z, \tz) f(g)\, , \nn \\
     && \cK_g(z, \tz) = \sum\limits_{k \in \N} \frac{\sqrt{k+1}}{k!} \brar{\tz}\epsilon g^{-1} \ket{z}^k \, . \nn 
\ee
When applied to Wigner matrix elements in the coherent state basis this map has an interesting interpretation: it essentially (up to some combinatorial factors) restricts the representation matrices of $\SU(2)$, when written in terms of spinors, to their holomorphic part
\be 
 D^j_{\omega \tilde{\omega}}(g) = \left( \bra{\omega} \frac{\ket{z}\brar{\tz} - \ketr{z}\bra{\tz} }{\sqrt{\braaket{z}{z} \braaket{\tz}{\tz}}}  \ket{\tilde{\omega}}  \right)^{2j} \stackrel{\cT}{\mapsto} \frac{1}{(2j)!\sqrt{d_j}}\braaket{\omega}{z}^{2j}\brar{\tz} \epsilon \ket{\tilde{\omega}}^{2j} \, . \nn
\ee
The unitary map (\ref{map}) directly generalizes from a single edge $e$ to an arbitrary graph $\Gamma$, showing unitary equivalence between the Hilbert spaces $\cH_\Gamma$ and $\cH^{\rm spin}_\Gamma = \otimes_e \cH_e^{\rm spin}$:
\be 
\cT_\Gamma: \cH_\Gamma \rightarrow \cH^{\rm spin}_\Gamma \, . \nn
\ee
Thus, equivalence classes of spinor functions living on different graphs $\Gamma$ and $\Gamma'$ can be defined by demanding the following diagram to commute
\be 
\xymatrix{
  \cH_\Gamma \ar[r]^{\cT_{\Gamma}} \ar[d]_{\,^*p_{\Gamma \Gamma'}} & **[r] \cH^{\rm spin}_{\Gamma}  \ar[d]^{\,^*p^{\rm spin}_{\Gamma \Gamma'}}
  \\
  \cH_{\Gamma'}    \ar[r]^{\cT_{\Gamma'}}   & **[r] \cH^{\rm spin}_{\Gamma'}
} \nn
\ee
Here $^*p_{\Gamma \Gamma'}$ are the isometric embeddings that define equivalence classes on the group side. Their counterparts on the spinor side $^*p_{\Gamma \Gamma'}^{\rm spin}$ are then used to define equivalence classes of spinor states. Thus, equivalence classes on the left side by construction are mapped to equivalence classes on the right side, no matter which $\cT_\Gamma$ is used.  This assures that the construction is cylindrically consistent and allows to abstractly define the \emph{continuum spinor Hilbert space} as
\be
\cH_{\rm LQG}^{\rm spin} := \overline{ \cup_\Gamma \cH_\Gamma^{\rm spin} / \sim  } \, . \nn 
\ee
Although the exact properties of this space are, for the moment, not very well understood, this shows that the spinor tools can be lifted from a fixed graph to the continuum level.
The Gaussian form of the measure, together with the simple polynomial form of the holomorphic basis (\ref{basis}), is expected to lead to simplification for practical computations: quantities of interest concern the moments of a simple Gaussian measure on $\C^4$ for which combinatorial tools, such as Wick's theorem, are available.

\section{Twistors and covariant twisted geometries} \label{sec:cov_twisted}

The construction described in section \ref{sec:twisted_geo} can be extended to $\SL(2,\C)$, and gives a notion of covariant twisted geometries based on the Lorentz group and not just SU(2). This is relevant for spin foam models and for the projected spin networks used in covariant versions of loop quantum gravity.
In all these models the starting point is the 12d phase space $T^*\SL(2,\C) \simeq \SL(2,\C) \times \Sl(2,\C)$, which is associated to each edge. Analogous to the construction of $T^*\SU(2)$ from spinors, discussed in section \ref{sec:twisted_geo}, one can construct the cotangent bundle $T^*\SL(2,\C)$ from \emph{twistors}
\cite{Dupuis:2011wy, Livine:2011vk,Wieland:2011ru}. 
While the algebra can be represented on $\T$ alone, representing the group element requires a second twistor. To avoid confusion with the previous twistorial description of $T^*\SU(2)$, we denote this time $\ket{t}$ and $\ket{u}$ the right- and left-handed components of $Z(u,t)$. We take the same invariant symplectic structure,
\be \label{ut1}
\{ \ket{t} , \bra{u} \} = -i\, \id.
\ee
We associate this twistor to the initial node of each edge, and a partner 
$\tilde Z(\tilde u,\tilde t)$ to the final node, and equipped with the same brackets  \eqref{ut1}. As before, we must eliminate some degenerate configurations from our description, which are now the cases $\braaket{u}{t}=0$ and $\braaket{\tilde u}{\tilde t}=0$. We denote by $\T_{2\ast}$ the non-degenerate space. We then consider \cite{Livine:2011vk}
\be \label{Group_element}
\vec{J}{^L} = \frac{1}{2}\braaaket{t}{\vec{\sigma}}{u}, & &  \qquad \vec{J}^R = \frac{1}{2}\braaaket{u}{\vec{\sigma}}{t}, 
\qquad G = \frac{\ket{t}\brar{\tilde{t}} - \ketr{u}\bra{\tilde{u}}}{\sqrt{\braaket{u}{t}  \braaket{\tilde{u}}{\tilde{t}}}}, \nn \\
\vec{\tilde{J}}^L = \frac{1}{2}\braaaket{\tilde{t}}{\vec{\sigma}}{\tilde{u}}, & & \qquad \vec{\tilde{J}}^R = \frac{1}{2}\braaaket{\tilde{u}}{\vec{\sigma}}{\tilde{t}},
\ee
and a similar area matching condition as before,
\be\label{cM1}
\cM = \braaket{u}{t} - \braaket{\tilde{u}}{\tilde{t}},
\ee
which ensures that both twistors on a given edge have the same (complex) helicity, but not necessarily vanishing as before.
This time the constraint is complex, and its real and imaginary part form a first class system, whose gauge transformations are the $\U(1)^\C$ transformations
\be
\ket{t} \mapsto e^{+\frac{i}{2}\beta}\ket{t}, \quad \ket{u} \mapsto  e^{+\frac{i}{2}\bar{\beta}}\ket{u}, \quad \ket{\tilde{t}} \mapsto e^{-\frac{i}{2}\beta}\ket{\tilde{t}}, \quad \ket{\tilde{u}} \mapsto  e^{-\frac{i}{2}\bar{\beta}}\ket{\tilde{u}}, \quad \beta \in \C,
\ee
which leaves \eqref{Group_element} invariant.
Hence, the symplectic reduction removes four dimensions. On the reduced 12d surface $\T_\ast \times \T_\ast/\!/\cM$,
a lengthy but simple computation shows that the coordinates \eqref{Group_element} satisfy the Poisson-algebra of $T^*\SL(2,\C)$, with $J$ and $\tilde J$ again right- and left-invariant vector fields, and $G$ in the defining right-handed representation $\bf{(0,1/2)}$. By taking the hermitian conjugate $G^\dagger$, or alternatively by exchanging the spinors for their duals and vice versa in (\ref{Group_element}), one gets a left-handed representation $\bf{(1/2,0)}$.

This leads to a generalization of theorem \ref{thm1} to the Lorentzian case,
\begin{theorem} \label{thm3} \cite{Livine:2011vk}
The symplectic reduction of the space $\T_\ast \times \T_\ast$ by the constraint $\cM$ is isomorphic to the cotangent bundle $T^*\SL(2,\C) - \{ |J| = 0  \}$ as a symplectic space.
\end{theorem}
 
A reduction to the previous $\SU(2)$ case is obtained if we identify the canonical SU(2) subgroup of unitary matrices, via $G^\dagger=G^{-1}$ and $\vec{J}{^L} = \vec{J}{^R}.$
This is achieved if we set $\ket{u}=\ket{t}:=\ket{z}$ and $\ket{\tilde u}=\ket{\tilde t}:=\ket{\tz}$. Then, \eqref{Group_element} reduce to (\ref{vector},\ref{g}), and the area matching \eqref{cM1} to \eqref{cM}. 

On the complete graph, we have a twistor per half-edge, and adapt the notation to $Z^n=(u^n,t^n)$, as before. The constraints are the complex area matching conditions \eqref{cM1} on each edge, and the $\SL(2,\C)$ closure condition on each node, which can be written in the chiral decomposition as $\sum_{e\in n} \vec{J}^R_e=\sum_{e\in n} \vec{J}^L_e={0}$.
In the same manner as \eqref{spinornet}, the structure of such a \emph{twistor network} is summarized by an action principle
\be \label{actiontwi}
\hspace*{-.4cm} \cS_\Gamma[t_e^{s,t}, u_e^{s,t}]\equiv \int d\tau \sum_e  -i\la u_e^{s,t}|\partial_\tau t_e^{s,t} \ra-i\la t_e^{s,t}|\partial_\tau u_e^{s,t} \ra 
+ \Phi_e (\la u_e^s|t_e^s\ra-\la u_e^t| t_e^t\ra)  + \sum_n \sum_{e \ni n} \la t_e^n|\Theta_n|u_e^n\ra
\ee
where the complex scalars $\Phi_e$ and the complex traceless matrices $\Theta_n$ are Lagrangian multipliers.

A geometric interpretation of the twistor networks is obtained doubling up the SU(2) picture of a collection of polyhedra. 
We now have a pair of spinors, $(u_e,t_e)$, for each face around a node $n$, and accordingly 
a bivector $J^{IJ}=(\vec{J}^L,\vec{J}^R)$ via \eqref{Group_element}. The bivector represents the two-normal to the face embedded in Minkowski spacetime, in the frame of $n$. 
The chiral closure conditions together with Minkowski's theorem imply the existence of \emph{two} polyhedra,
corresponding to the right- and left-handed sectors. A Hopf section decomposition similar to \eqref{twigeo} can be also given, see \cite{Livine:2011vk} for details.
The geometric interpretation becomes more interesting if one includes the simplicity constraints. As we review below, this amounts to identifying the right- and left-handed polyhedra,
and leads to a notion of \emph{covariant} twisted geometries, a collection of 3d polyhedra with arbitrary $\SL(2,\C)$ curvature among them.

\subsection{$GL(N,\C)$ formalism}

Before moving on, let us briefly discuss the algebra of invariants, which will play an important role in the following. 
As for the $\SU(2)$ case, a basis of invariants is given by the scalar products among bivectors, but these fail to form a proper algebra, again because of \eqref{JJJ}
which is still valid on each right/left sector. A solution to this problem can be found using spinors. As shown in \cite{Dupuis:2011wy}, 
an (over)complete basis for the space of global $\SL(2,\C)$ invariants is given by the following quantities,
\be
\la u_i|t_j \ra\,=\,A_{ij}-iB_{ij},
\quad
\la t_i|u_j \ra\,=\,A_{ij}+iB_{ij},
\quad
[t_i|t_j \ra\,=\,F_{ij}-iG_{ij},
\quad
[u_i|u_j \ra\,=\,F_{ij}+iG_{ij}.
\ee
The matrices $A_{ij}$ and $B_{ij}$ are Hermitian, while $F_{ij}$ and $G_{ij}$ are antisymmetric and holomorphic.
The invariants form a genuine algebra \cite{Dupuis:2011wy}, of which $A$ and $B$ generate a $\mathfrak{gl}(N,\C)$ subalgebra, which is the complexification of the $\mathfrak{u}(N)$ algebra found in the case of $\SU(2)$  invariants. 

\section{Holomorphic simplicity constraints} \label{sec:simplicity}

In covariant formulations of Loop Quantum Gravity, such as spin foam models, a crucial role is played by the simplicity constraints.
In the continuum theory, they guarantee that the bi-vectors $B^{IJ}$ define a unique tetrad.\footnote{More precisely, they imply that
the two tetrads naturally defined by $B^{IJ}$ are to be identified, see \cite{bimetric}.} At the discrete level, they can be realized as quadratic equations on the space of $\SL(2,\C)$ invariants,
\be \label{simplquadra}
\vec{J}^{R}_e\cdot\vec{J}^{R}_f - e^{2i\theta} \,\vec{J}^{L}_e\cdot\vec{J}^{L}_f = 0,  \qquad \theta = 2\arctan \gamma,
\ee
where $\gamma$ is the Immirzi parameter. Alternatively, as linear equations on each edge,
$\vec{K}_e+\gamma \vec{J}_e = 0$ (in the time gauge). The linear version is the one used in the EPRL spin foam models, and can be also rewritten as
\be \label{simpllin}
\vec{J}^R_e + e^{i\theta} \vec{J}^L_e =0.
\ee
The constraints are second class, and do not form an algebra on the gauge-invariant intertwiner space, both aspects being direct consequences of \eqref{JJJ}.
This poses notorious difficulties at the quantum level, which have been the focal point of various discussions in the literature. 

The spinorial formalism offers a new and powerful way to approach this issue. In fact, one can find a version of the constraints, quadratic in the spinors, which implies \eqref{simplquadra} and \eqref{simpllin}, and which does form a genuine algebra on the space of $\SL(2,\C)$ invariants. It is based on the invariants discussed earlier.
The new simplicity constraints have been introduced in \cite{Dupuis:2010iq} for the Euclidean case, and in \cite{Dupuis:2011wy} for the Lorentzian case. 
They take the form of an antisymmetric matrix,
\be\label{simple_def}
\cC_{ef} \equiv [t_e|t_f \ra - e^{i\theta} [u_e|u_f \ra = 0.
\ee
The constraints are holomorphic with respect to the natural complex structure of the spinors, and furthermore they Poisson commute with each other:
\be
\{\cC_{ef}, \cC_{gh} \} =0,
\ee
while of course $\{\cC_{ef}, \overline \cC_{gh} \} \neq 0.$ 
This is the key property of such holomorphic simplicity constraints, which has important applications at the quantum level.
Notice that because of the Pl\"ucker relations, there are only $2N-3$ independent constraints per node. Nevertheless, they can all be imposed harmlessly since they commute. See \cite{Dupuis:2011wy} for more details. Observe also that the distinction between diagonal and off-diagonal constraints, familiar from the quadratic version \eqref{simplquadra}, now disappears, with the advantage of a proper algebra and a clear holomorphic factorization.

If we add the holomorphic simplicity constraints \eqref{simple_def} to the action of twistor networks \eqref{actiontwi}, we get a notion of \emph{simple} twistor networks, 
\be\label{actionsimple}
\cS^{simple}_\Gamma[t^{s,t}_e,u^{s,t}_e]
=\cS_\Gamma[t^{s,t}_e,u^{s,t}_e] + \int d\tau\, \sum_n\sum_{e,f\ni n} \Psi_{e,f} \Big([t^n_e|t^n_f\ra - e^{i\theta} [u^n_e|u^n_f\ra\Big),
\ee
with $\Psi_{e,f} $ a suitable Lagrange multiplier.
The role of the simplicity constraints is then to identify the right- and left-handed sectors as in \eqref{simpllin}, up to a $\gamma$-dependent phase.
This identifies a unique polyhedron around each node, with the face bivectors all lying in the same 3d spacelike surface, plus a timelike normal, $N_n^I$, encoded in the spinors \cite{Dupuis:2011wy}. The role of the Immirzi parameter is to determine the true area bivector as $B^{IJ}=(\id -\gamma\, \star) J^{IJ}$.
This information can be effectively traded for a \emph{single} spinor $ \ket{z^n_e} $ per half-edge, and one pure boost $\Lambda_n \in \SL(2,\C)/\SU(2)$ per node, such that $\ket{t^n_e} = \Lambda_n \ket{z^n_e}$ and  $\ket{u^n_e} = (\Lambda_n)^{-1} \ket{z^n_e}$ \cite{Dupuis:2011wy}.
In other words, a simple twistor network describes a covariant twisted geometry: a collection of closed 3d polyhedra with arbitrary $\SL(2,\C)$ curvature among them.

These simple twistor networks are very interesting from the perspective that they contain the same information as a normal spinor network for $\SU(2)$, but allow to describe its natural embedding into a $\SL(2,\C)$-invariant structure, through the introduction of non-trivial time-normals living at each vertex of the graph $\Gamma$.
They provide a classical version of the simple projected spin networks \cite{Dupuis:2010jn}, which form the boundary Hilbert space of EPRL/FK spin foam models \cite{Dupuis:2010jn,Rovelli:2010ed}.
Furthermore, the existence of a new, holomorphic formulation of the simplicity constraints \eqref{simple_def}, allows a new treatment of the quantization, in which all the constraints are treated on the same footing and imposed strongly thanks to their commutativity. This program has been realized in 
\cite{Dupuis:2010iq, Dupuis:2011fz,Dupuis:2011dh}  for the Euclidean case, where it was shown its equivalence to a weak imposition, \'a la Gupta-Bleuler, of the original second class quadratic constraints. Exact solutions have been constructed as coherent $\U(N)$ intertwiners, and the resulting spinfoam amplitudes 
are directly written as a discrete action in terms of spinors and holonomies. For the Lorentzian case, the construction of suitable coherent states has not appeared yet, and the quantization program is in progress \cite{DupuisInprogress}.

\section{Summary and Outlook}

Describing the classical phase space of loop quantum gravity in terms of spinor variables appears to be a powerful idea. The standard holonomy-flux structure is derived from a much simpler collection of spinors on a graph, a spinor network. This is linked to a discrete geometric picture, the twisted geometries \cite{Freidel:2010aq,Freidel:2010bw}. The spinors give a different Hilbert space representation for loop gravity and new calculational tools \cite{Livine:2011gp, Livine:2011zz}, 
which are expected to find many application: physical quantities such as correlation functions in loop quantum gravity involve complicated $\SU(2)$ integrals, 
which with the help of spinors become Gaussian integrals over the complex plane, a priori much easier to handle.
Furthermore, these ideas extend naturally to covariant descriptions in which the full Lorentz group appears.
Using the relation to twistors \cite{Freidel:2010bw}, it is straighforward to extend the spinorial formalism to  $\SL(2,\C)$, described in terms of \emph{twistor networks} \cite{Livine:2011vk,Wieland:2011ru}. This allows the identification of new holomorphic simplicity constraints \cite{Dupuis:2011wy}, classically equivalent to the usual linear and quadratic constraints, but with the key property that they Poisson commute. Including these constraints, one gets simple twistor networks describing covariant twisted geometries,
a collection of 3d polyhedra with arbitrary $\SL(2,\C)$ curvature among them.

Finally, for some references, recent developments using spinor techniques in the context of loop gravity and spin foams include a toy-model for quantum cosmology \cite{Borja:2010gn, Borja:2010hn, Borja:2011yk, Livine:2011up}, a new look on the simplicity constraints in spinfoam models \cite{Dupuis:2010iq, Dupuis:2011fz, Dupuis:2011dh, Dupuis:2011wy},
advances in the context of group field theory \cite{Dupuis:2011fx} and topological BF-theory \cite{Bonzom:2011nv}. This spinorial framework in the covariant context could provide a new angle of attack  to tackle the issue of renormalization and coarse-graining of spin foam amplitudes and group field theories \cite{Dupuis:2011dh, Dupuis:2011fx}.

\section*{Acknowledgements}
Simone would like to thank the organizers of the School for the invitation to lecture. This work was partially supported by the ANR ``Programme Blanc'' grant LQG-09. 
\medskip\\

\bibliographystyle{JHEP}

\begin{thebibliography}{10}

\bibitem{Rovelli:2010km}
C.~Rovelli and S.~Speziale, {\it {On the geometry of loop quantum gravity on a
  graph}},  {\em Phys.Rev.} {\bf D82} (2010) 044018
  [\href{http://arXiv.org/abs/1005.2927}{{\tt 1005.2927}}].

\bibitem{Ashtekar:1998ak}
A.~Ashtekar, A.~Corichi and J.~A. Zapata, {\it {Quantum theory of geometry III:
  Noncommutativity of Riemannian structures}},  {\em Class.Quant.Grav.} {\bf
  15} (1998) 2955--2972 [\href{http://arXiv.org/abs/gr-qc/9806041}{{\tt
  gr-qc/9806041}}].

\bibitem{Thiemann:2000bv}
T.~Thiemann, {\it {Quantum spin dynamics (QSD): 7. Symplectic structures and
  continuum lattice formulations of gauge field theories}},  {\em
  Class.Quant.Grav.} {\bf 18} (2001) 3293--3338
  [\href{http://arXiv.org/abs/hep-th/0005232}{{\tt hep-th/0005232}}].

\bibitem{Bianchi:2009tj}
E.~Bianchi, {\it {Loop Quantum Gravity a la Aharonov-Bohm}},
  \href{http://arXiv.org/abs/0907.4388}{{\tt 0907.4388}}.

\bibitem{Freidel:2011ue}
L.~Freidel, M.~Geiller and J.~Ziprick, {\it {Continuous formulation of the Loop
  Quantum Gravity phase space}},  \href{http://arXiv.org/abs/1110.4833}{{\tt
  1110.4833}}.

\bibitem{Baratin:2010nn}
A.~Baratin, B.~Dittrich, D.~Oriti and J.~Tambornino, {\it {Non-commutative flux
  representation for loop quantum gravity}},  {\em Class.Quant.Grav.} {\bf 28}
  (2011) 175011 [\href{http://arXiv.org/abs/1004.3450}{{\tt 1004.3450}}].

\bibitem{Freidel:2010aq}
L.~Freidel and S.~Speziale, {\it {Twisted geometries: A geometric
  parametrisation of SU(2) phase space}},  {\em Phys.Rev.} {\bf D82} (2010)
  084040 [\href{http://arXiv.org/abs/1001.2748}{{\tt 1001.2748}}].

\bibitem{Freidel:2010bw}
L.~Freidel and S.~Speziale, {\it {From twistors to twisted geometries}},  {\em
  Phys.Rev.} {\bf D82} (2010) 084041
  [\href{http://arXiv.org/abs/1006.0199}{{\tt 1006.0199}}].

\bibitem{Bianchi:2010gc}
E.~Bianchi, P.~Dona and S.~Speziale, {\it {Polyhedra in loop quantum gravity}},
   {\em Phys.Rev.} {\bf D83} (2011) 044035
  [\href{http://arXiv.org/abs/1009.3402}{{\tt 1009.3402}}].

\bibitem{Freidel:2009ck}
L.~Freidel and E.~R. Livine, {\it {The Fine Structure of SU(2) Intertwiners
  from U(N) Representations}},  {\em J.Math.Phys.} {\bf 51} (2010) 082502
  [\href{http://arXiv.org/abs/0911.3553}{{\tt 0911.3553}}].

\bibitem{Freidel:2010tt}
L.~Freidel and E.~R. Livine, {\it {U(N) Coherent States for Loop Quantum
  Gravity}},  {\em J.Math.Phys.} {\bf 52} (2011) 052502
  [\href{http://arXiv.org/abs/1005.2090}{{\tt 1005.2090}}].

\bibitem{Borja:2010rc}
E.~F. Borja, L.~Freidel, I.~Garay and E.~R. Livine, {\it {U(N) tools for Loop
  Quantum Gravity: The Return of the Spinor}},  {\em Class.Quant.Grav.} {\bf
  28} (2011) 055005 [\href{http://arXiv.org/abs/1010.5451}{{\tt 1010.5451}}].

\bibitem{Borja:2011pd}
  E.~F.~Borja, J.~Diaz-Polo and I.~Garay,
  ``U(N) and holomorphic methods for LQG and Spin Foams,''
  arXiv:1110.4578 [gr-qc].
  
\bibitem{Dupuis:2011wy}
M.~Dupuis, L.~Freidel, E.~R. Livine and S.~Speziale, {\it {Holomorphic
  Lorentzian Simplicity Constraints}},
  \href{http://arXiv.org/abs/1107.5274}{{\tt 1107.5274}}.

\bibitem{Bahr:2007xn}
B.~Bahr and T.~Thiemann, {\it {Gauge-invariant coherent states for loop quantum
  gravity. II. Non-Abelian gauge groups}},  {\em Class.Quant.Grav.} {\bf 26}
  (2009) 045012 [\href{http://arXiv.org/abs/0709.4636}{{\tt 0709.4636}}].

\bibitem{Dittrich:2008va}
B.~Dittrich and S.~Speziale, {\it {Area-angle variables for general
  relativity}},  {\em New J.Phys.} {\bf 10} (2008) 083006
  [\href{http://arXiv.org/abs/0802.0864}{{\tt 0802.0864}}].

\bibitem{Dittrich:2008ar}
B.~Dittrich and J.~P. Ryan, {\it {Phase space descriptions for simplicial 4d
  geometries}},  {\em Class.Quant.Grav.} {\bf 28} (2011) 065006
  [\href{http://arXiv.org/abs/0807.2806}{{\tt 0807.2806}}].

\bibitem{Girelli:2005ii}
F.~Girelli and E.~R. Livine, {\it {Reconstructing quantum geometry from quantum
  information: Spin networks as harmonic oscillators}},  {\em
  Class.Quant.Grav.} {\bf 22} (2005) 3295--3314
  [\href{http://arXiv.org/abs/gr-qc/0501075}{{\tt gr-qc/0501075}}].

\bibitem{Livine:2011gp}
E.~R. Livine and J.~Tambornino, {\it {Spinor Representation for Loop Quantum
  Gravity}},  \href{http://arXiv.org/abs/1105.3385}{{\tt 1105.3385}}.

\bibitem{Livine:2011zz}
E.~R. Livine and J.~Tambornino, {\it {Loop gravity in terms of spinors}},
  \href{http://arXiv.org/abs/1109.3572}{{\tt 1109.3572}}.

\bibitem{QTwiGeo}
L.~Freidel, E.~Livine and S.~Speziale, {\it {Quantum twisted geometries and
  coherent states}}, . to appear.

\bibitem{Livine:2011vk}
E.~R. Livine, S.~Speziale and J.~Tambornino, {\it {Twistor Networks and
  Covariant Twisted Geometries}},  \href{http://arXiv.org/abs/1108.0369}{{\tt
  1108.0369}}.

\bibitem{Wieland:2011ru}
W.~M. Wieland, {\it {Twistorial phase space for complex Ashtekar variables}},
  \href{http://arXiv.org/abs/1107.5002}{{\tt 1107.5002}}.

\bibitem{bimetric}
S.~Speziale, {\it {Bi-metric theory of gravity from the non-chiral Plebanski
  action}},  {\em Phys.Rev.} {\bf D82} (2010) 064003
  [\href{http://arXiv.org/abs/1003.4701}{{\tt 1003.4701}}].

\bibitem{Dupuis:2010iq}
M.~Dupuis and E.~R. Livine, {\it {Revisiting the Simplicity Constraints and
  Coherent Intertwiners}},  {\em Class.Quant.Grav.} {\bf 28} (2011) 085001
  [\href{http://arXiv.org/abs/1006.5666}{{\tt 1006.5666}}].

\bibitem{Dupuis:2010jn}
M.~Dupuis and E.~R. Livine, {\it {Lifting SU(2) Spin Networks to Projected Spin
  Networks}},  {\em Phys.Rev.} {\bf D82} (2010) 064044
  [\href{http://arXiv.org/abs/1008.4093}{{\tt 1008.4093}}].

\bibitem{Rovelli:2010ed}
C.~Rovelli and S.~Speziale, {\it {Lorentz covariance of loop quantum gravity}},
   {\em Phys.Rev.} {\bf D83} (2011) 104029
  [\href{http://arXiv.org/abs/1012.1739}{{\tt 1012.1739}}].

\bibitem{Dupuis:2011fz}
M.~Dupuis and E.~R. Livine, {\it {Holomorphic Simplicity Constraints for 4d
  Spinfoam Models}},  \href{http://arXiv.org/abs/1104.3683}{{\tt 1104.3683}}.

\bibitem{Dupuis:2011dh}
M.~Dupuis and E.~R. Livine, {\it {Holomorphic Simplicity Constraints for 4d
  Riemannian Spinfoam Models}},  \href{http://arXiv.org/abs/1111.1125}{{\tt
  1111.1125}}.

\bibitem{DupuisInprogress}
M.~Dupuis, L.~Freidel, E.~R. Livine and S.~Speziale, {\it {SL(2,C) Intertwiners
  and Simplicity for Spinfoam Models}}, . To appear.

\bibitem{Borja:2010gn}
E.~F. Borja, J.~Diaz-Polo, I.~Garay and E.~R. Livine, {\it {Dynamics for a
  2-vertex Quantum Gravity Model}},  {\em Class.Quant.Grav.} {\bf 27} (2010)
  235010 [\href{http://arXiv.org/abs/1006.2451}{{\tt 1006.2451}}].

\bibitem{Borja:2010hn}
E.~F. Borja, J.~Diaz-Polo, I.~Garay and E.~R. Livine, {\it {U(N) invariant
  dynamics for a simplified Loop Quantum Gravity model}},
  \href{http://arXiv.org/abs/1012.3832}{{\tt 1012.3832}}.

\bibitem{Borja:2011yk}
E.~F. Borja, J.~Diaz-Polo, L.~Freidel, I.~Garay and E.~R. Livine, {\it
  {Dynamics for a simple graph using the U(N) framework for loop quantum
  gravity}},  \href{http://arXiv.org/abs/1110.6017}{{\tt 1110.6017}}.

\bibitem{Livine:2011up}
E.~R. Livine and M.~Martin-Benito, {\it {Classical Setting and Effective
  Dynamics for Spinfoam Cosmology}},
  \href{http://arXiv.org/abs/1111.2867}{{\tt 1111.2867}}.

\bibitem{Dupuis:2011fx}
M.~Dupuis, F.~Girelli and E.~R. Livine, {\it {Spinors and Voros star-product
  for Group Field Theory: First Contact}},
  \href{http://arXiv.org/abs/1107.5693}{{\tt 1107.5693}}.

\bibitem{Bonzom:2011nv}
V.~Bonzom and E.~R. Livine, {\it {A New Hamiltonian for the Topological BF
  phase with spinor networks}},  \href{http://arXiv.org/abs/1110.3272}{{\tt
  1110.3272}}.

\end{thebibliography}
\providecommand{\href}[2]{#2}\begingroup\raggedright\endgroup

\end{document}